\documentclass[journal=jpccck,manuscript=article]{achemso}
\setkeys{acs}{usetitle=true}
\usepackage[version=3]{mhchem} 
\usepackage{multirow}
\usepackage{natmove}
\usepackage{bm}				
\author{Caterina Cocchi}
\email{caterina.cocchi@unimore.it}
\affiliation{Centro S3, CNR-Istituto Nanoscienze, I-41125 Modena, Italy}
\altaffiliation{Present address: Humboldt-Universit\"at zu Berlin, Institut f\"ur Physik und IRIS Adlershof, Zum Grossen Windkanal 6, 12489 Berlin, Germany; e-mail: caterina.cocchi@physik.hu-berlin.de}
\author{Deborah Prezzi}
\affiliation{Centro S3, CNR-Istituto Nanoscienze, I-41125 Modena, Italy}
\author{Alice Ruini}
\affiliation{Centro S3, CNR-Istituto Nanoscienze, I-41125 Modena, Italy}
\alsoaffiliation{Dipartimento di Scienze Fisiche, Informatiche e Matematiche, Universit\`a di Modena e Reggio Emilia, I-41125 Modena, Italy}
\author{Marilia J. Caldas}
\affiliation{Instituto de F{\'\i}sica, Universidade de S\~ao Paulo, 05508-900 S\~ao Paulo, SP, Brazil}
\author{Annalisa Fasolino}
\affiliation{Institute for Molecules and Materials, Radboud University Nijmegen,
 Heyendaalseweg 135, 6525AJ~Nijmegen, The Netherlands}
\author{Elisa Molinari}
\affiliation{Centro S3, CNR-Istituto Nanoscienze, I-41125 Modena, Italy}
\alsoaffiliation{Dipartimento di Scienze Fisiche, Informatiche e Matematiche, Universit\`a di Modena e Reggio Emilia, I-41125 Modena, Italy}
\title{Concavity Effects on the Optical Properties of Aromatic Hydrocarbons}
\begin{document}
\begin{abstract}
We study the modifications on the ground and excited state properties of polycyclic aromatic hydrocarbons (PAHs), induced by the variation of concavity and $\pi$-connectivity. Inspired by experimentally feasible systems, we study three series of PAHs, from H-saturated graphene flakes to geodesic buckybowls, related to the formation of fullerene \ce{C60} and \ce{C50}-carbon nanotube caps.
Working within the framework of quantum chemistry semi-empirical methods AM1 and ZINDO/S, we find that the interplay between concavity and $\pi$-connectivity shifts the bright optical lines to higher energies, and introduces symmetry-forbidden dark excitations at low energy.
A generally good agreement with the available experimental data supports our results, which can be viewed as the basis for designing optical properties of novel curved aromatic molecules.
\end{abstract}
\newpage
The discovery of fullerene \cite{krot+91cr} and carbon nanotubes (CNTs) \cite{iiji91nat} and, more recently, of graphene \cite{novo+04sci}, each with unique physical and chemical properties \cite{dres+96book,kats12book}, has promoted a continuous research on carbon nanoscience \cite{hohe+10ac}.
In this context, non planar polycyclic aromatic hydrocarbons (PAHs), including buckybowls and geodesic polyarenes, have gained increasing importance both as synthesis intermediates \cite{boor+01sci,yu+10nl,amsh+10ac,scot+12jacs,tsef-scot06cr} and as intriguing systems themselves \cite{wu-sieg06cr,kawa-kura06cr,jack+07jacs,mack+07obc,stei+09jacs,whal+11cs}.
Indeed the interest in identifying the path bringing flat aromatic molecules into fullerene cages extends also to other fields than organic chemistry, as for instance to astrochemistry\cite{bern-tiel12pnas}.
In another direction, crystal-packed structures \cite{fork+97jacs,xiao+08jacs} and supramolecular compounds made of curved PAHs \cite{pere-mart08csr} can be nowadays synthesized with promising perspectives towards molecular nanodevices \cite{sygu+07jacs,wu+08jacs,bald+10obc,trem+10chpch,zopp+11jacs,osel+12nano}.

Since the main features of carbon-based materials are known to be critically influenced by morphological details, including edge modifications \cite{prez+08prb,jian+08prl,li+10jacs,zhu-su10jpcc,wang+10jpca,cocc+11jpcc,cocc+11jpcl,prez+11prb,wang-wang12jpcc} and structural distortions \cite{caet+09lang,wong09jpcc,rieg-muel10jpoc,cocc+12jpcc}, we here address the effects, induced by variations of concavity and $\pi$-connectivity, on the ground and excited state properties of PAHs.
Inspired by the synthesis intermediates of \ce{C60}-fullerene and \ce{C50}-CNT-caps \cite{amsh+10ac,scot+12jacs,jack+07jacs,chuv+10natc}, we study three series of symmetric PAHs, ranging from planar molecules to concave buckybowls, and compare our results with the available theoretical and experimental data \cite{jack+07jacs,scot+12jacs,bend+91ijqc,ajie+90jpc}.
The ground state properties of these PAHs are characterized by the appearance of a net dipole in the out-of-plane direction, as soon as the molecule starts to deviate from planarity.
The optical absorption is also very sensitive to the increase of both concavity and $\pi$-connectivity, which produces a blue shift of the bright optical lines, and at the same time introduces symmetry-forbidden dark excitations at low energy.
These features can represent the basis for engineering the optical properties of novel curved aromatic molecules.

%
\section{Methods and Systems}
The presented results have been obtained within the framework of Hartree-Fock-based semi-empirical methods.
All the structures have been optimized through the AM1 model \cite{dewa+85jacs} (0.4 $\text{kcal} \cdot \text{mol}^{-1}$/\AA{} threshold for the forces), which demonstrated its reliability for the investigation of curved aromatics \cite{dina-nara02jmst,petr+05joc}.
The optical spectra are evaluated by means of ZINDO/S \cite{ridl-zern73tca}, which implements single-excitation Configuration Interaction (CIS),
\bibnote{AM1 and ZINDO/S calculations were performed using VAMP package included in Accelrys Materials Studio software, version 6.0  (\url{http://accelrys.com/products/materials-studio}).
Our convergence tests over a number of occupied and virtual molecular orbitals (MOs) indicated that a CI energy window of at least 4.5 eV below the HOMO and 3.5 eV above the LUMO is required for a reliable characterization of the low energy optical excitations.}
and has been validated for \ce{C60} in the pioneering work by Bendale \textit{et al.} \cite{bend+91ijqc}.
To highlight the dipolar character of the optical excitations, we introduce the transition density (TD), defined as:
\begin{equation}
\rho^I (\mathbf{r}) = \sum_{a,r} c^I_{ar} \phi_{r}^* (\mathbf{r}) \phi_{a} (\mathbf{r}) ,
\label{TD}
\end{equation}
where $c^I_{ar}$ are the CI coefficients of the $I^{th}$ excited state, corresponding to single excitations from the occupied $\phi_{a}$ to the virtual molecular orbitals $\phi_{r}$.
The sign distribution of this quantity provides a clear pictorial view of the excitation dipole.

\begin{figure}
\centering
\includegraphics[width=.95\textwidth]{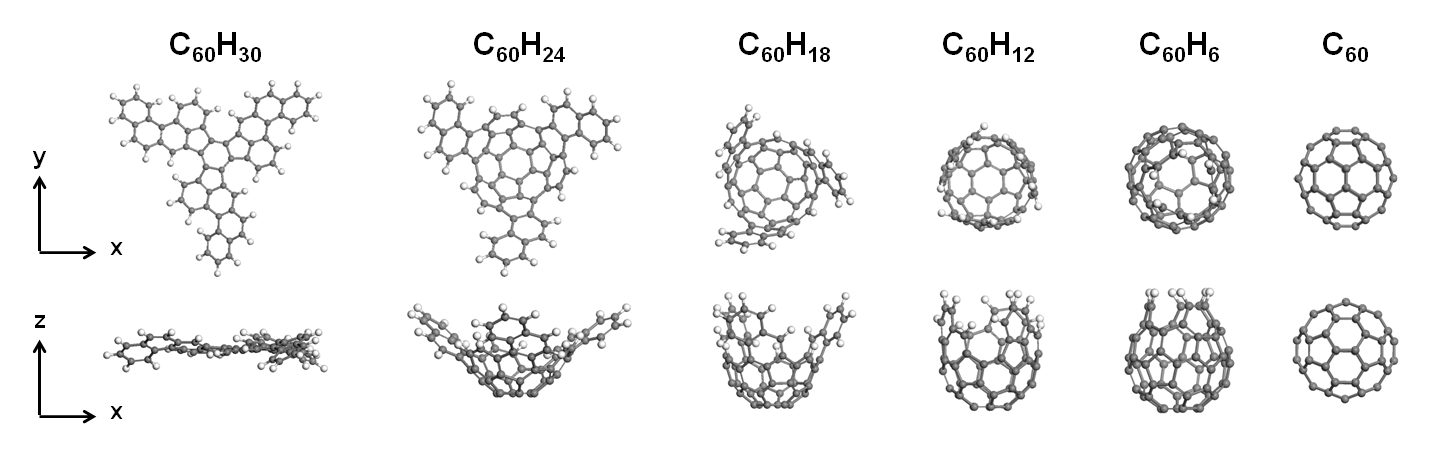}%
\caption{Top ($x$,$y$) and side ($x$,$z$) view of the \ce{C60H_{6n}} ($n = 0,... ,5$) series of PAHs, related to the formation of icosahedral \ce{C60} ($I_h$ point group symmetry). At each step, three pairs of H atoms are removed, promoting the $\pi$-connectivity of C atoms at the edges and inducing the concavity. The $C_3$ symmetry of the initial structure \ce{C60H30} is preserved up to the last intermediate \ce{C60H6}.
}
\label{fig1}
\end{figure}

In this work we focus on three series of isolated PAHs in gas phase, presenting increasing concavity.
In \ref{fig1} we show the series of \ce{C60H_{6n}} ($n=0, ..., 5$) PAHs, experimentally identified as precursors for an efficient bottom-up synthesis of \ce{C60}-fullerene catalyzed on a metal surface \cite{amsh+10ac}.
These structures contain only hexagons and pentagons and are characterized by a $C_3$ symmetry, that leads naturally to the 20 hexagons and 12 pentagons of the \ce{C60} with icosahedral symmetry ($I_h$ point group), upon step-by-step removal of three pairs of H atoms at the periphery.
In the process of cage formation the molecules undergo a variation of both concavity and $\pi$-connectivity.
Concavity is induced by the closure of the network already tailored in the planar \ce{C60H30}, with the progressive formation of pentagons and hexagons.
On the other hand, $\pi$-connectivity can be quantified as the number of $\pi$-bonds formed by each C atom in the molecule.
The $\pi$-connectivity index rises from 2.5 in \ce{C60H30}, where 30 C atoms out of 60 are H-terminated, up to 3 in the \ce{C60} molecule, where only C-bonds with $sp^2$ character are present.

\section{Results and Discussion}
In the \ce{C60} series, depicted in \ref{fig1}, we first notice that a net dipole is acquired by the molecules as soon as the concavity starts to develop: the magnitude of this $z$-oriented dipole moment is related to the interplay between the extension of the molecule in the  out-of-plane direction and the presence of a H-passivated edge.
Starting from a negligible intrinsic dipole of the initial planar structure \ce{C60H30} ($\mu_z$=0.16 D), which is to be ascribed to the slight distortions of the molecule (see \ref{fig1}), the dipole moment rapidly increases already for the first curved molecule \ce{C60H24} ($\mu_z$=7.25 D) and reaches a maximum for \ce{C60H18} ($\mu_z$=10.18 D), which presents a $\pi$-connectivity index equal to 2.7.
As the $\pi$-connectivity further increases and the cage progressively closes, the magnitude of the dipole moment starts to decrease ($\mu_z$=9.71 D and 6.22 D for \ce{C60H12} and \ce{C60H6}, respectively), eventually vanishing for \ce{C60}.
The process of cage closure described above is also accompanied by an increase of both ionization potential and electron affinity from \ce{C60H30} to \ce{C60}.
This trend witnesses the increasing stabilization and reduced reactivity from the planar H-terminated structure, to the fullerene cage.
For further details, see Supporting Information, Table S1.
\begin{figure}
\centering
\includegraphics[width=.45\textwidth]{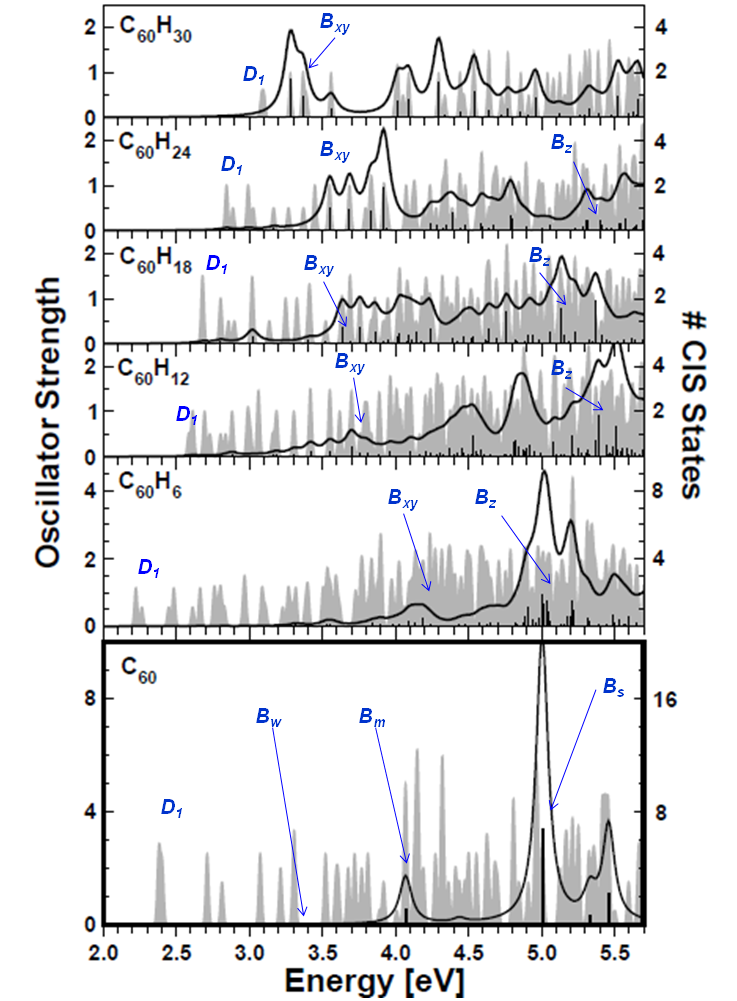}%
\caption{Optical spectra of the \ce{C60H_{6n}} series: the excitations are indicated by solid bars and the curves are obtained through a Lorentzian broadening with full width at half maximum of 100 meV.
The density of CIS states shown in the background (grey shaded area) is obtained by counting each excited state with a Gaussian broadening with standard deviation of 20 meV.
In each panel the lowest symmetry-forbidden excitation is labelled as $D_1$: in some spectra more excitations with different symmetry appear summed up to $D_1$ in the density of CIS states.
Bright peaks are indicated as $B_{xy}$ and $B_z$, depending on their polarization, along the reference directions reported in \ref{fig1}.
In the spectrum of \ce{C60} (bottom panel), bright excitations, which are all triply degenerate, are identified according to their intensity, as in Ref. \cite{bend+91ijqc}: $B_w$ is referred to the weakest peak, $B_m$ to a medium-intensity peak and $B_s$ to the strongest excitation.
}
\label{fig2}
\end{figure}

\begin{figure}
\centering
\includegraphics[width=.95\textwidth]{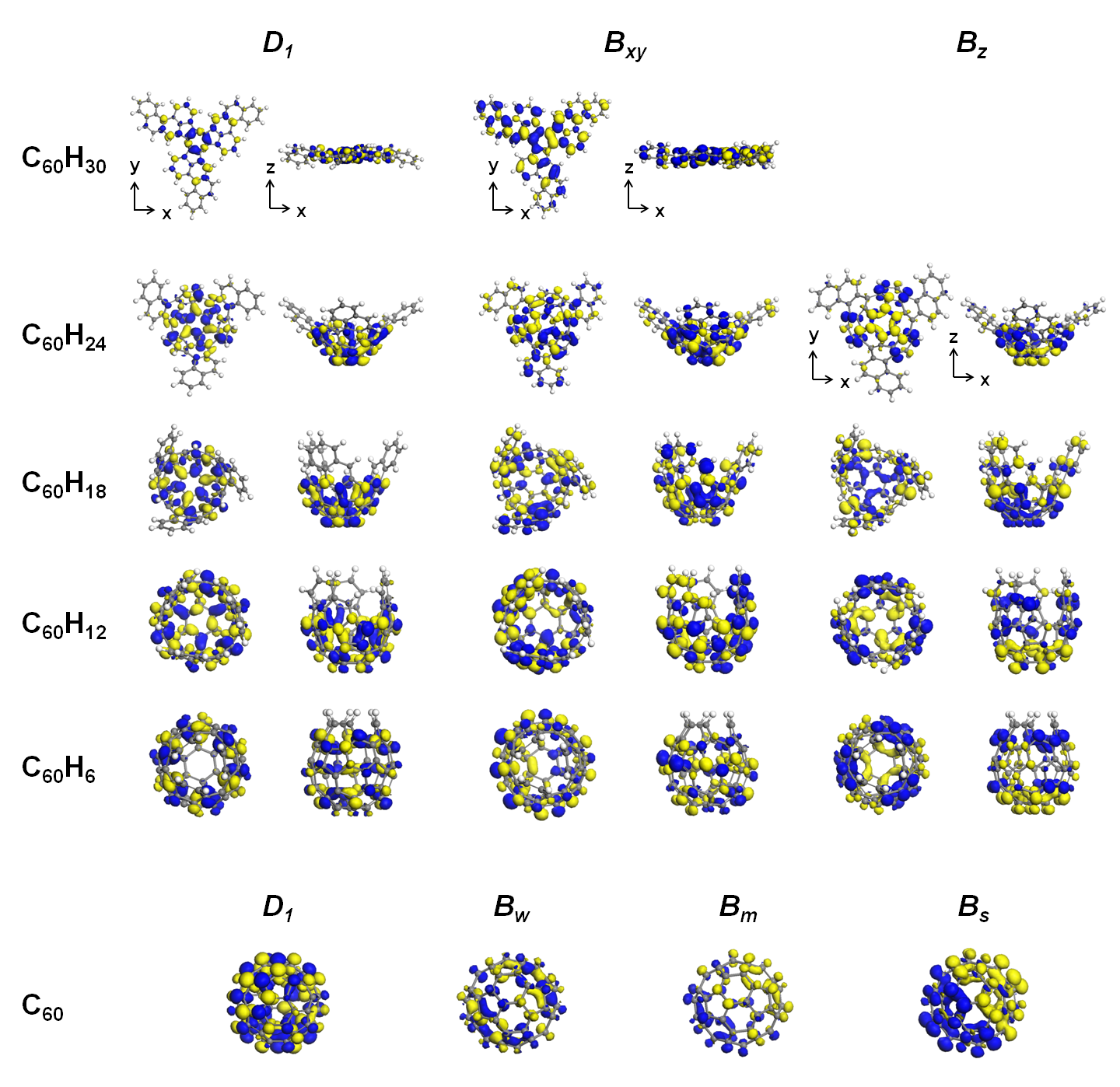}%
\caption{Top ($x,y$) and side ($x,z$) view of the transition density plots of the excitations of the \ce{C60H_{6n}} series ($n = 0,... ,5$), as indicated in the spectra of \ref{fig2}.
}
\label{fig3}
\end{figure}

In \ref{fig2} we present the optical spectra of the \ce{C60H_{6n}} PAH series: the excitations are indicated by solid bars, broadened by a Lorentzian curve with a full width at half maximum (FWHM) of 100 meV.
In the background of each panel we also report the density of CIS states (shaded grey curve), obtained by counting each excitation, irrespective of its oscillator strength (OS), and by applying for visualization a Gaussian broadening with standard deviation of 20 meV.
Starting from the spectrum of \ce{C60H30}, the lowest energy excitations appear already at about 3.1 eV: we indicate as $D_1$ the first symmetry-forbidden excitation (see \ref{table1} and  \ref{fig3}).
The first bright peak, labelled as $B_{xy}$, according to its polarization along the $x$ and $y$ directions, appears at 3.3 eV. 
This excitation is doubly degenerate, and its strong intensity is given by the large dipole moment in the ($x$,$y$) plane (see the transition density plot in \ref{fig3}).
Due to the planar structure of \ce{C60H30}, the peaks at higher energy are also ($x,y$)-polarized.

\begin{table*}
\begin{tabular}{c|c c|c c|c c c}
\multirow{2}{*}{\textbf{Structure}}  & \multicolumn{2}{|c|}{\textbf{$D_1$}}  & \multicolumn{2}{|c|}{\textbf{$B_{xy}$}} &  \multicolumn{2}{|c}{\textbf{$B_z$}}  \\ \cline{2-7}
 & \textbf{Energy [eV]} & \textbf{OS} &  \textbf{Energy [eV]} & \textbf{OS} & \textbf{Energy [eV]} & \textbf{OS}  \\ \hline
\ce{C60H30} & 3.08 (s) & 0.00 & 3.28 (d) & 0.84 & - & - \\
\ce{C60H24} & 2.89 (s) & 0.00 & 3.55 (d) & 0.50 & 5.54 (s) & 0.16  \\
\ce{C60H18} & 2.68 (s) & 0.00 & 3.63 (d) & 0.37 & 5.13 (s) & 0.76 \\
\ce{C60H12} & 2.59 (s) & 0.00 &  3.70 (d) & 0.21 &  5.39 (s) & 0.90  \\
\ce{C60H6} & 2.26 (s) & 0.00 & 3.90 (d) & 0.01 &  5.00 (s) & 0.94  \\ \hline
\ce{C60} & 2.38 (t) & 0.00 & 3.30 (t)  & 0.0007 ($B_w$) & 5.01 (t) & 3.40 ($B_s$)  \\
 & & & 4.07 (t) & 0.56 ($B_m$) & & \\
\end{tabular}
\caption{Energy and oscillator strength (OS) of the main excitations of the PAH series \ce{C60H_{6n}} and of \ce{C60}, including the first symmetry-forbidden excitation ($D_1$) and the bright peaks polarized in the ($x$,$y$)-plane ($B_{xy}$) and along the $z$-direction ($B_z$).
In the case of \ce{C60}, we adopt the notation of Ref. \cite{bend+91ijqc} to label the bright peaks.
While in \ce{C60} all the reported excitations present triple (t) degeneracy, in the H-terminated PAH of this series the indicated excitations present either single (s) or double (d) degeneracy.
}
\label{table1}
\end{table*}

\begin{figure}
\centering
\includegraphics[width=.95\textwidth]{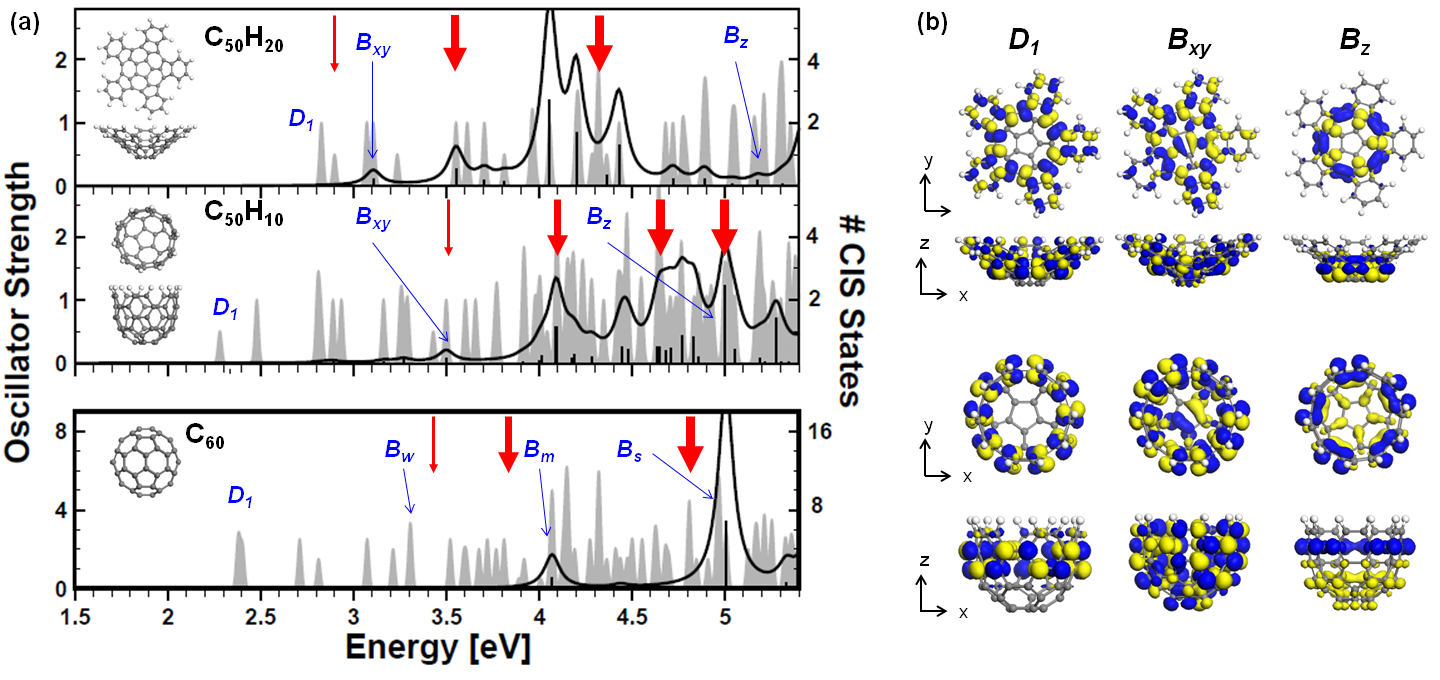}%
\caption{(a) Optical spectra of pentaindenocorannulene (\ce{C50H20}) and of the related CNT-cap (\ce{C50H10}), shown in the inset; for comparison, the spectrum of \ce{C60} is displayed on the bottom panel.
In analogy with the \ce{C60H_{6n}} series, shown in \ref{fig2}, also in the spectra of \ce{C50H20} and \ce{C50H10} the lowest dark excitation ($D_1$) as well as the first bright peaks ($B_{xy}$ and $B_z$) are marked.
The excitations are indicated by solid bars and the curves are obtained through a Lorentzian curve with full width at half maximum of 100 meV.
The grey shaded lines represent the density of CIS states (Gaussian broadening with standard deviation of 20 meV).
Red arrows point to the energy of optical peaks observed experimentally \cite{jack+07jacs,scot+12jacs,bend+91ijqc} for \ce{C50H20}, \ce{C50H10} and \ce{C60}, respectively.
The thickness of the arrows indicates of the intensity of the detected excitations.
(b) Top ($x$,$y$) and side ($x$,$z$) views of the transition densities associated with $D_1$, $B_{xy}$ and $B_z$ for \ce{C50H20} (top) and \ce{C50H10} (bottom).
}
\label{fig4}
\end{figure}

We can compare the features of \ce{C60H30} to those of \ce{C60}, which represents the opposite extreme of the series. 
Starting with the symmetry-forbidden transitions, in addition to the lowest energy dark excitation ($D_1$), which is triply degenerate due to the $I_h$ symmetry of fullerene and is now at about 2.3 eV (see \ref{table1} and \ref{fig2}), many other dark excitations appear in the low energy region.
For further details in this regard see Table S3 and the related discussion in the Supporting Information).
The allowed absorption peaks are shifted in this case to considerably higher energies: notably, three bright peaks are identified and labelled according to the notation of Bendale \textit{et al.} \cite{bend+91ijqc}, where \textit{weak}, \textit{medium} and \textit{intense} excitations are mentioned below 5.5 eV.
These peaks are indicated in the spectrum of \ce{C60} in \ref{fig2} as $B_w$, $B_m$ and $B_s$, respectively, and  they are also triply degenerate (see \ref{table1}). 
$B_w$ is a very weak excitation (OS < $10^{-3}$), which is observed at 3.3 eV.
A more intense peak ($B_m$) is visible in the spectrum at about 4.1 eV, while the strongest absorption peak ($B_s$) is found at 5 eV.
The sign modulation of the transition density of $B_w$ clearly explains the low intensity of this excitation (see \ref{fig3}).
On the contrary, the transition densities associated with $B_m$ and $B_s$ present separate domains of opposite sign.
For a given value of isosurface (8 $\times$ $10^{-4}$ in this case), the higher intensity of $B_s$ compared to $B_m$ can be visualized in the plots in \ref{fig3}.
Comparing our results with those presented by Bendale and coworkers \cite{bend+91ijqc}, we notice a good agreement with the excitation energies obtained for a comparable CI window.
\bibnote{Our calculations were performed including 38 occupied and 37 empty states, which correspond to an energy window of 6.9 eV below HOMO and 5.5 eV above LUMO.
This CI window is compatible with that including 30 occupied and 35 empty states in Ref. \cite{bend+91ijqc} and indeed the two sets of results are consistent (see \ref{table1} for detailed comparison).
According to the results of Bendale \textit{et al.}, only $B_w$ and $B_m$ are convergent within a threshold of 100 meV with respect to the largest employed CI window, with 111 occupied and 109 empty states \cite{bend+91ijqc}.
The strongest peak at about 5 eV undergoes a shift of over 300 meV between the 30 $\times$ 35 and 111 $\times$ 109 CI windows \cite{bend+91ijqc}.
Further slight discrepancies between our results and those obtained in Ref. \cite{bend+91ijqc} can be ascribed to the differently optimized geometries: AM1 is adopted here, while Bendale \textit{et al.} make use of the INDO/1 hamiltonian.
The resulting lengths of pentagon-hexagon and hexagon-hexagon bonds result 1.464 \AA{} and 1.398 \AA{}, which are 0.013 \AA{} larger and shorter, respectively, than those computed in Ref. \cite{bend+91ijqc}.}
A thorough comparison with the experimental data taken from Ref. \cite{ajie+90jpc} is presented in the discussion of \ref{fig4}.

Focusing now on the series behavior reported in \ref{fig2}, we observe a blue shift of the intense absorption lines and a red shift of the lowest energy dark state $D_1$, with the appearance of an increasingly large number of forbidden excitations in between.
These features are due to the increase of both concavity and $\pi$-connectivity, triggered by the dehydrogenation process in the formation of the fullerene cage.
The increase of $\pi$-connectivity leads to a larger delocalization of the molecular orbitals (MOs), which is responsible for the general red shift of the density of forbidden or very weak CIS states (see \ref{fig2}, from the top to the bottom panel).
\bibnote{In the \ce{C60H_{6n}} series $D_1$ is found energetically close (few tens of meV) to a pair of doubly degenerate excitations polarized in the ($x$,$y$) plane, with very weak but non zero intensity (OS $\sim$ $10^{-4}$ -- $10^{-2}$).
It is worth noting that this behavior finally leads to manifolds of triply degenerate excitations in \ce{C60}.
In the case of \ce{C60H24}, \ce{C60H18} and \ce{C60H6}, the excitation $D_1$ corresponds to the third excited state in the spectrum, as a pair of doubly degenerate excitation is found at lower energy.}

The first bright peak $B_{xy}$, observed for \ce{C60H30}, is easily recognized for \ce{C60H24} and for \ce{C60H18} as the first intense peak in the spectrum.
For \ce{C60H12} and \ce{C60H6} the identification of $B_{xy}$ is less immediate, since the low energy part of their spectra is characterized by a large number of low intensity peaks polarized in the ($x$,$y$) plane.
In the spectra of these molecules $B_{xy}$ is chosen for the similarity of the transition density with respect to the less concave structures \ce{C60H24} and \ce{C60H18} (see the plots in \ref{fig3} and \ref{table1} for the trends of energy and OS).
On the other hand, the intensity of the absorption increases for more concave molecules at higher energy.
In this region the most intense peaks present a $z$-polarization.
$B_z$ is identified in the spectra as the first $z$-polarized peak: the orientation of its excitation dipole is evident in the transition density plots in \ref{fig3}.
A remarkable similarity between the spectra of \ce{C60H6} and \ce{C60} is noticed, suggesting that the intense $B_z$ peak in the former molecule eventually becomes the strong $B_s$ peak in \ce{C60}.

The optical features discussed for the \ce{C60H_{6n}} series are apparent also for another set of PAHs, namely pentaindenocorannulene (\ce{C50H20}) and the [5,5]-CNT-cap (\ce{C50H10}).
Both these structures have been experimentally synthesized in the last few years, as crystal-packed structures in solution \cite{jack+07jacs,stei+09jacs,scot+12jacs}.
Differently from the structures of the \ce{C60H_{6n}} series, these systems have not been produced as subsequent synthesis steps, although clear perspectives in this sense are suggested in the outlook of Ref. \cite{stei+09jacs}.
Both pentaindenocorannulene and the CNT-cap belong to $C_{5v}$ point group symmetry and are characterized by the same central corannulene unit (see insets in \ref{fig4}a), which represents the starting point in the process of synthesis of both \ce{C50H20} \cite{jack+07jacs} and \ce{C50H10} \cite{scot+12jacs}.
Due to the similarities between these structures and the increasing $\pi$-connectivity and concavity from \ce{C50H20} to \ce{C50H10}, it is reasonable to carry out an analytical comparison of their optical properties, in analogy with the \ce{C60H_{6n}} series.
The trends observed for this set of PAHs are in agreement with the \ce{C60H_{6n}} series, concerning the electronic and optical properties.
Both \ce{C50H20} and \ce{C50H10} are characterized by a dipole moment in the out-of-plane $z$-direction ($\mu_z$=2.59 D and 8.98 D, respectively).
In analogy with the \ce{C60H_{6n}} series, we observe increasing ionization potential and electron affinity from \ce{C50H20} to the CNT-cap, as a consequence of larger electronic stability at higher $\pi$-connectivity index (2.6 for \ce{C50H20} and 2.8 for \ce{C50H10}).
Further details are provided in the Supporting Information, Table S2.

The optical spectra of these molecules are shown in \ref{fig4}a, in comparison with the spectrum of the closed-cage \ce{C60}, reported for reference.
As for the case of the \ce{C60H_{6n}} series, we identify in the spectra the excitations labelled as $D_1$, $B_{xy}$ and $B_z$, in addition to a number of dark states lying between $D_1$ and the first active peak.
The energy of the first excitation decreases from \ce{C50H20} to the CNT-cap, on account of the increased $\pi$-connectivity.
On the other hand, the energy of bright peaks blue shifts in the spectrum of the cap.
$B_{xy}$ is identified in the spectra of both \ce{C50H20} and \ce{C50H10} as the lowest energy doubly degenerate peak polarized in the ($x$,$y$) plane (see the transition densities in \ref{fig4}b).
The intensity of $B_z$ grows by one order of magnitude from \ce{C50H20} to \ce{C50H10}, due to the more extended C-conjugated network along the $z$ direction in the CNT-cap.
For further details, see the Supporting Information, Table S4.

The computed spectra of \ce{C50H20} and \ce{C50H10} are compared in \ref{fig4}a with the available experimental data presented in the Supporting Information of Refs. \cite{jack+07jacs, scot+12jacs}, respectively.
In both cases the spectra are recorded in \ce{CH2Cl2} solution.
The experimentally detected energy of the main peaks is indicated in \ref{fig4}a by the position of red arrows, and their intensity is represented by the thickness of the arrows.
By inspecting \ref{fig4}a, a generally good agreement between our results and the available experimental data is shown, concerning both the energy and the strength of the intense peaks.
Our study represents the first theoretical investigation on the optical properties of pentaindenocorannulene \ce{C50H20} and of the CNT-cap \ce{C50H10} and we are hopeful that it can further stimulate the interest in the interaction of these molecules with light, also from the experimental viewpoint.

In the bottom panel of \ref{fig4}a we report the optical spectrum of \ce{C60}, in comparison with the experimental UV-vis spectrum of Ref. \cite{ajie+90jpc}, as discussed also in Ref. \cite{bend+91ijqc}.
In our calculations, the \textit{medium} ($B_m$) and \textit{strong} ($B_s$) peaks at 4 and 5 eV result almost systematically blue shifted of about 200 meV compared to the experimental spectra, taken in hexane solution \cite{ajie+90jpc}.
An even better agreement is obtained for the \textit{weak} peak $B_w$, which is detected at 3.4 eV and predicted by our calculations to be at 3.3 eV.
It is worth noting that the lowest energy excitations, which are indicated as dark by symmetry in our ZINDO/S spectrum, are actually observed in the experimental spectra.
This testifies that the manifold of dark excitations in the low energy region of the spectra of concave PAHs can be activated in experiments, due to a loss of symmetry of the molecules, as it likely occurs in solution or under other laboratory conditions.

\begin{figure}
\centering
\includegraphics[width=.45\textwidth]{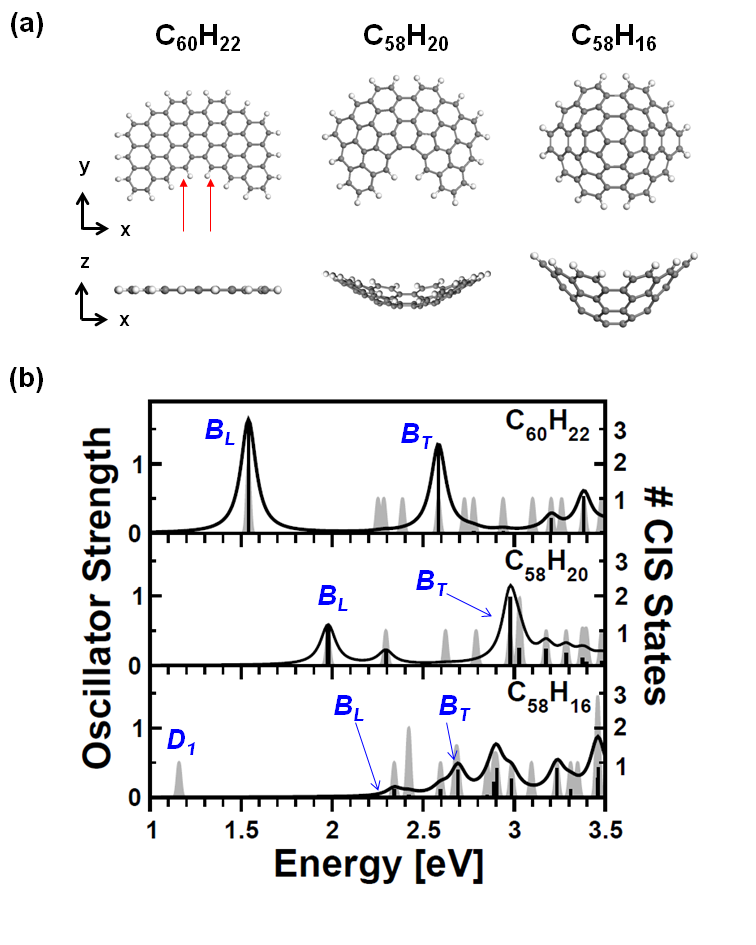}%
\caption{(a) Top ($x$,$y$) and side view ($x$,$z$) of planar (\ce{C60H22}), etched (\ce{C58H20}) and zipped (\ce{C58H16}) PAHs.
The structure of \ce{C58H20} is obtained by removing the two C atoms, and the corresponding saturating H atoms, in the cove region of \ce{C60H20} (see red arrows).
Two pentagons are formed in \ce{C60H22} and next, after the removal of 4 H atoms at the edges, the flake is zipped through the bonding of the unsaturated C atoms.
(b) Optical spectra of the systems shown in (a): the excitations are indicated by solid bars and the curves are obtained by adding a Lorentzian broadening with FWHM of 100 meV.
The density of CIS states (Gaussian broadening with standard deviation of 20 meV) is indicated in the background of each panel, by the grey shaded area.
In the spectrum of the planar \ce{C60H22} the first bright peaks are labelled as $B_L$ and $B_T$, according to their longitudinal (along $x$ axis) and transverse (along $z$ axis) polarization, respectively.
The corresponding excitations are indicated also in the spectra of the etched \ce{C58H20} and zipped \ce{C58H16}.
}
\label{fig5}
\end{figure}

As a last example, we discuss a third class of PAHs, where the five-member rings are not included in the initial structure, but tend to form in defective graphene flakes, as shown in \ref{fig5}a.
These structures are inspired by those presented in Ref. \cite{chuv+10natc}.
In that work, the formation of \ce{C60} from a graphene sheet is visualized in real time through a transmission electron microscope.
The process is modelled in four steps: the initial edge etching of a graphene flake with over 80 C atoms is followed by the formation of pentagons and then by the curving of the graphene into a bowl-shape structure.
Finally, the open edges are zipped up to form a closed fullerene structure.
We here model the etching and zipping of graphene samples using H-passivated graphene flakes with about 60 C atoms, for comparison with the \ce{C60H_{6n}} series of fullerene precursors.
Starting from an initially planar structure (\ce{C60H22}, \ref{fig5}a), two C atoms are removed by etching and two pentagons are formed in the \ce{C58H20} intermediate molecule, introducing concavity.
The molecule is finally zipped after the removal of 4 H atoms at the cove edge and the subsequent bonding of the unsaturated C atoms.
No additional pentagons are formed in this process and the resulting \ce{C58H16} assumes a bowl-like shape.
The initial structure \ce{C60H22} and the zipped flake \ce{C58H16} present $C_{2v}$ symmetry, which are however related to different rotation axes.
While \ce{C60H22} is symmetric for rotation with respect to the $y$ axis, the symmetry axis of \ce{C58H16} is in the $z$ direction (see \ref{fig5}a).

The optical spectra of this set of structures are shown in \ref{fig5}b.
For the planar \ce{C60H22} the low energy region of the spectrum is characterized by two intense peaks polarized in the longitudinal ($x$) and transverse ($y$) direction of the molecules ($B_L$ and $B_T$, respectively), similarly to the situation of H-terminated armchair graphene nanoribbons with finite length \cite{cocc+12jpcl}.
$B_L$ and $B_T$ are identified also in the spectrum of \ce{C58H20} and their polarization directions are preserved even after the formation of pentagons.
An overall blue shift of the spectrum of \ce{C58H20} is observed, compared to that of \ce{C60H22}: it is induced by the increased concavity acquired by the molecule after the etching.
This behavior confirms the picture drawn for the \ce{C60H_{6n}} series.
The intensity of $B_L$, dominated by a HOMO$\rightarrow$LUMO transition, decreases as the extension of \ce{C58H20} in the $x$ direction is reduced; on the other hand the intensity of $B_T$ is basically preserved from \ce{C60H22} to \ce{C58H20}.
The edge zipping from \ce{C58H20} to \ce{C58H16} contributes to the increase of both the concavity and the $\pi$-connectivity, whose index raises from about 2.66 to about 2.72.
As a consequence of the induced concavity, the first bright peak $B_L$ is blue shifted (see \ref{fig5}b) and its OS is reduced, owing to the smaller extension of the flake along the polarization direction $x$.
As an effect of $\pi$-connectivity, we note an interesting behavior of the occupied MOs, which leads to the appearance in the spectrum of \ce{C58H16} of a symmetry forbidden state ($D_1$) at very low energy.
This excitation is due to the HOMO$\rightarrow$LUMO transition, but in this case the HOMO has the same character of the HOMO-3 of \ce{C58H20}, which moves here to higher energy with respect to the other two MOs, due to the increased $\pi$-connectivity.

\section{Conclusions}
In conclusion, we have examined three families of experimentally feasible PAHs, related to the formation of concave C-allotropes (including \ce{C60} and \ce{C50}-CNT-caps), having different underlying symmetry and increasing concavity and $\pi$-connectivity.
The interplay between these properties is responsible for a general blue shift of the optical absorption and for the appearance of a number of dark excitations at low energy, specifically in the visible region of the spectrum, which are forbidden by symmetry.
The high energy optical modes polarized in the out-of-plane direction gain oscillator strength, due to the increasing concavity and to the presence of a H-passivated edge.
On the other hand, the excitations polarized in the ($x$,$y$) plane, which dominate the spectra of the planar structures, are considerably less intense. 
At the same time, as expected, the increase of $\pi$-connectivity tends to reduce the energy of the first excited configurations.
The resulting lowest-energy states are dark, but they may however be experimentally detected, due to overall symmetry-loosening of the structure.

The trends observed in our results are mainly based on general symmetry and topology arguments and can therefore represent the basis for engineering the optical properties of novel curved aromatic molecules.
These features are observed in all the considered PAH series and are supported by a good agreement with the available experimental data.
For most of the considered systems, the results presented here are the first theoretical study on the optical properties.
We believe that our detailed analysis on the excitations, with specific attention devoted to the dark states in the visible range, will further stimulate the interest in curved PAHs and specifically in the their interaction with light.

\begin{acknowledgement}
The authors are grateful to Lennert van Tilburg for useful discussions.
This work was partly supported by the Italian Ministry of University and Research
under FIRB grant ItalNanoNet, and by Fondazione Cassa di Risparmio di Modena with
project COLDandFEW.
M.~J.~C.~ acknowledges support from FAPESP and CNPq (Brazil).
\end{acknowledgement}
\begin{suppinfo}
We include in the Supporting Information the computed electron affinity and ionization potential for the \ce{C60H_{6n}} series and the details of the optical excitations (i.e. excitation energies and oscillator strength as well as transition density plots) of \ce{C60}, \ce{C50H20}and \ce{C50H10} and zipped PAHs.
\end{suppinfo}
\providecommand*{\mcitethebibliography}{\thebibliography}
\csname @ifundefined\endcsname{endmcitethebibliography}
{\let\endmcitethebibliography\endthebibliography}{}

\begin{tocentry}
\includegraphics[width=9cm]{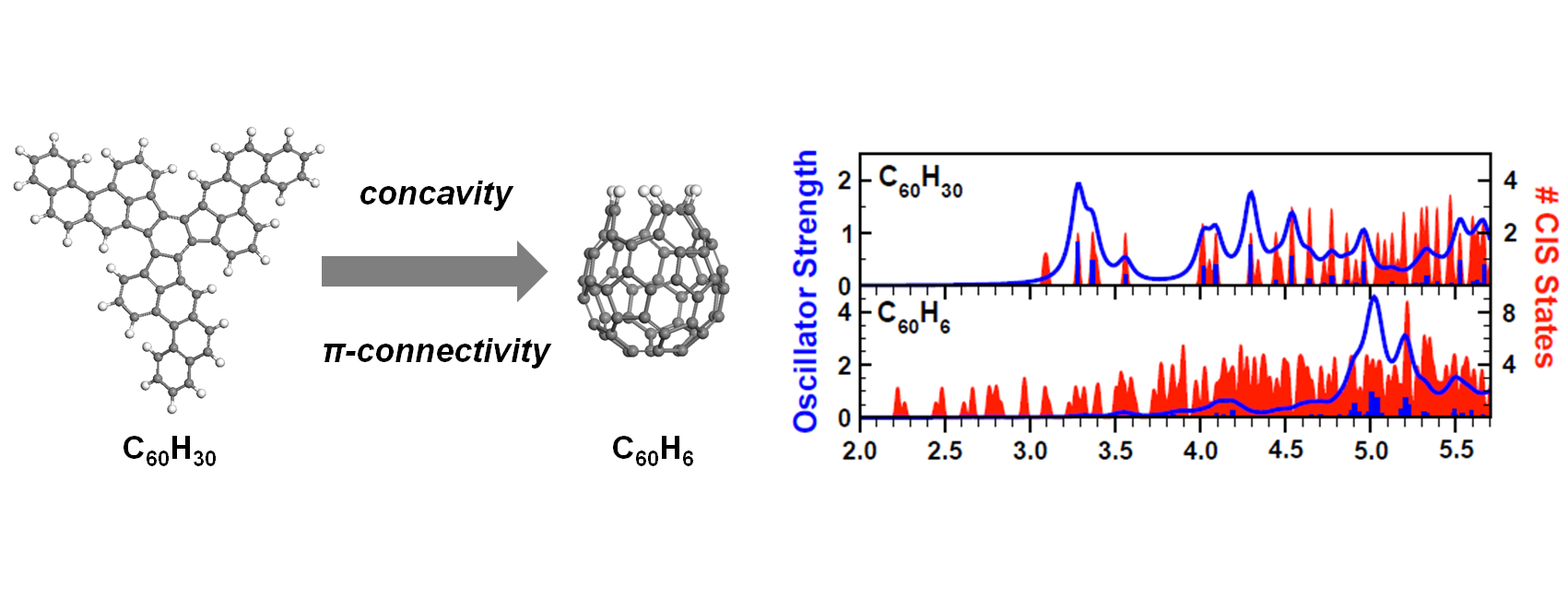}
\end{tocentry}

\begin{mcitethebibliography}{52}
\bibitem[Not()]{Note-3}
 In the \ce{C60H_{6n}} series $D_1$ is found energetically close (few tens of
  meV) to a pair of doubly degenerate excitations polarized in the ($x$,$y$)
  plane, with very weak but non zero intensity (OS $\sim$ $10^{-4}$ --
  $10^{-2}$). It is worth noting that this behavior finally leads to manifolds
  of triply degenerate excitations in \ce{C60}. In the case of \ce{C60H24},
  \ce{C60H18} and \ce{C60H6}, the excitation $D_1$ corresponds to the third
  excited state in the spectrum, as a pair of doubly degenerate excitation is
  found at lower energy.\relax
\mciteBstWouldAddEndPunctfalse
\mciteSetBstMidEndSepPunct{\mcitedefaultmidpunct}
{}{\mcitedefaultseppunct}\relax
\EndOfBibitem
\end{mcitethebibliography}
\end{document}